\documentstyle[prl,aps,floats,epsf,psfig]{revtex}

\def\nim#1#2#3  {{ Nucl. Instr. Meth.} {\bf#1}, #2 (#3). }
\def\np#1#2#3   {{ Nucl. Phys.} {\bf#1}, #2 (#3). }
\def\pcps#1#2#3 {{ Proc. Cam. Phil. Soc.} {\bf#1}, #2 (#3). }
\def\pl#1#2#3   {{ Phys. Lett.} {\bf#1}, #2 (#3). }
\def\plc#1#2#3   {{ Phys. Lett.} {\bf#1}, #2 (#3). }
\def\prep#1#2#3 {{ Phys. Rep.} {\bf#1}, #2 (#3). }
\def\prev#1#2#3 {{ Phys. Rev.} {\bf#1}, #2 (#3). }
\def\prl#1#2#3  {{ Phys. Rev. Lett.} {\bf#1}, #2 (#3). }
\def\prs#1#2#3  {{ Proc. Roy. Soc.} {\bf#1}, #2 (#3). }
\def\ptp#1#2#3  {{ Prog. Th. Phys.} {\bf#1}, #2 (#3). }
\def\rmp#1#2#3  {{ Rev. Mod. Phys.} {\bf#1}, #2 (#3). }
\def\rpp#1#2#3  {{ Rep. Prog. Phys.} {\bf#1}, #2 (#3). }
\def\zp#1#2#3   {{ Z. Phys.} {\bf#1}, #2 (#3). }
\def\epj#1#2#3   {{ Eur. Phys. Jour.} {\bf#1}, #2 (#3). }

\begin{document}

\wideabs{
\title{Search for the Lepton Number Violating Process $\overline{\nu}_{\mu} e^- \rightarrow \mu^- \overline{\nu}_e$}
\author{
 J.~A.~Formaggio$^2$,
 J.~Yu$^3$, 
 T.~Adams$^{4}$,
 A.~Alton$^{4}$,
 S.~Avvakumov$^8$,
 L.~de~Barbaro$^5$,
 P.~de~Barbaro$^8$,
 R.~H.~Bernstein$^3$,
 A.~Bodek$^8$,
 T.~Bolton$^4$,
 J.~Brau$^6$,
 D.~Buchholz$^5$,
 H.~Budd$^8$,
 L.~Bugel$^{3}$,
 J.~M.~Conrad$^2$,
 R.~B.~Drucker$^6$,
 B.~T.~Fleming$^2$,
 J.~Foster$^4$,
 R.~Frey$^6$,
 J.~Goldman$^4$,
 M.~Goncharov$^4$,
 D.~A.~Harris$^3$,
 R.~A.~Johnson$^1$,
 J.~H.~Kim$^2$,
 S.~Koutsoliotas$^2$,
 M.~J.~Lamm$^3$,
 W.~Marsh$^3$,
 D.~Mason$^6$,
 J.~McDonald$^7$,
 K.~S.~McFarland$^8$,
 C.~McNulty$^2$,
 D.~Naples$^7$,
 P.~Nienaber$^{3}$,
 A.~Romosan$^2$,
 W.~K.~Sakumoto$^8$,
 H.~M.~Schellman$^5$,
 M.~H.~Shaevitz$^2$,
 P.~Spentzouris$^3$,
 E.~G.~Stern$^2$,
 N.~Suwonjandee$^1$,
 M.~Vakili$^1$,
 A.~Vaitaitis$^2$,
 U.~K.~Yang$^8$,
 G~.P.~Zeller$^5$, and  
 E.~D.~Zimmerman$^2$
}
\address{
$^1$ University of Cincinnati, Cincinnati, OH 45221 \\
$^2$ Columbia University, New York, NY 10027 \\
$^3$ Fermi National Accelerator Laboratory, Batavia, IL 60510 \\
$^4$ Kansas State University, Manhattan, KS 66506 \\
$^5$ Northwestern University, Evanston, IL 97403 \\
$^6$ University of Oregon, Eugene, OR 97403 \\
$^7$ University of Pittsburgh, Pittsburgh, PA 15260 \\
$^8$ University of Rochester, Rochester, NY 14627 \\
}
\date{\today}
\maketitle
\begin{abstract}
The NuTeV experiment at Fermilab has used a sign-selected neutrino
beam to perform a search for the lepton number violating process
$\overline{\nu}_{\mu} e^- \rightarrow \mu^-\overline{\nu}_e$, and to
measure the cross-section of the Standard Model inverse muon
decay process $\nu_{\mu} e^- \rightarrow \mu^- \nu_e$. NuTeV measures
the inverse muon decay asymptotic cross-section $\sigma/E$ to be
$(13.8 \pm 1.2 \pm 1.4) \times 10^{-42} $ cm$^2$/GeV.  The experiment
also observes no evidence for lepton number violation and places one
of the most restrictive limits on the cross-section ratio
$\sigma(\overline{\nu}_{\mu} e^- \rightarrow
\mu^-\overline{\nu}_e)/\sigma(\nu_{\mu} e^- \rightarrow \mu^- \nu_e)$
$\leq 1.7\% $ at 90\% C.L. for V$-$A couplings and $\leq 0.6\%$ for
scalar couplings.
\end{abstract}
\twocolumn
}

Neutrino-lepton interactions provide an excellent tool to study the
properties of the weak interaction. Such purely leptonic processes
experience no interference from strong coupling terms, and thus
provide a direct channel to investigate the nature of the weak force.  The
inverse muon decay (IMD) process:

\begin{equation}
\nu_{\mu} + e^- \rightarrow \mu^- + \nu_e
\end{equation}

\noindent allows one to make an accurate determination of the
vector/axial-vector (V$-$A) nature of the weak interaction
\cite{Fetcher}. This process is also sensitive to scalar couplings and
right-handed currents.

An experiment with separate neutrino and anti-neutrino beams can
search for the process:

\begin{equation}
\overline{\nu}_{\mu} + e^- \rightarrow \mu^- + \overline{\nu}_e
\end{equation}

\noindent Such an interaction is forbidden by the Standard Model,
since it violates lepton family number conservation ($\Delta L_e =
-\Delta L_{\mu} = 2$).  Theories which incorporate multiplicative
lepton number conservation \cite{Cabbibo}, left-right symmetry \cite{Herczeg}, or the existence of bileptons \cite{Cuypers}
allow for such processes to occur.

\begin{figure}
\centerline{
\psfig{figure=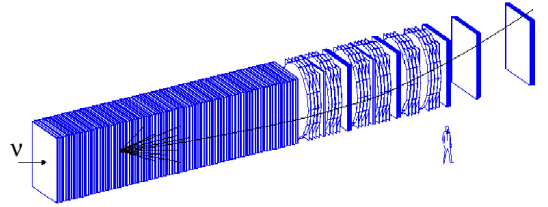,height=1.5in}}
\caption{Schematic of the NuTeV detector, showing the calorimeter and
the toroid spectrometer.}
\label{fig:labe}
\end{figure}

The NuTeV neutrino experiment at Fermilab has investigated
these processes in its high-energy, sign-selected neutrino
beamline.  Although the NuTeV inverse muon decay measurement is dominated by systematic uncertainties, the search for lepton number violation (LNV) processes is very sensitive because the relevant backgrounds are highly suppressed.

The experiment collected data during the 1996-1997 fixed target run,
receiving a total of $2.9 \times 10^{18}$ 800 GeV protons striking a
BeO target. Pions and kaons produced in the interaction were focused
using the Sign-Selected Quadrupole Train (SSQT)\cite{SSQT} and aimed toward the NuTeV detector at a 7.8 mrad angle relative to the primary proton beam
direction.  The SSQT enabled the detector to be exposed to either pure
neutrino or pure anti-neutrino beams. NuTeV received $1.3 \times
10^{18}$ and $1.6 \times 10^{18}$ protons on target for neutrino and
anti-neutrino running modes, respectively.  The fractional
contamination from wrong-sign meson decays was below $5
\times 10^{-3}$ \cite{nc_charm}. Pions and kaons decay to neutrinos as
they travel through a 440 m vacuum pipe; undecayed hadrons are
filtered out in a beam dump at the end of the pipe.  The neutrinos
pass through about 900 m of earth berm shielding before reaching the
NuTeV neutrino detector.

The NuTeV detector \cite{detector}, located 1.4 km downstream of the primary target,
consists of a segmented iron-scintillator sampling calorimeter,
followed by a toroid spectrometer (see Fig.~\ref{fig:labe}).  The
calorimeter is composed of 42 segments, each segment consisting of
four 2 inch thick steel plates, two liquid scintillator counters, and
one drift chamber.  The calorimeter serves as a
neutrino target with a fiducial mass of 350 tons.  The scintillation
counters measure the deposited hadronic energy and the drift chamber
determine the position and direction of the outgoing muon.  The toroid
spectrometer uses a 15 kG toroid magnetic field to measure the charge
and energy of muons exiting from the calorimeter.  The toroid magnetic field is configured so as to always focus muons coming from the selected neutrino beam ($\mu^-$ for neutrinos, $\mu^+$ for anti-neutrinos). The energy
resolution and response of the detector is measured directly using a
separate beam of hadrons, muons, and electrons at varying
energies. The hadronic energy resolution of the calorimeter is
$\sigma/E=(0.024\pm 0.001)\oplus(0.874\pm0.003)/\sqrt{E}$, and the
electromagnetic energy resolution is ${\sigma }/{E} =(0.04\pm
0.001)\oplus{(0.52\pm 0.01)}/{\sqrt{E}}$\cite{NIM}.  The resolution of
the muon energy as determined by the toroid spectrometer is $\Delta p
/ p = 11\%$, limited predominantly by multiple scattering.

The selection criteria for the inverse muon decay measurement and the
lepton number violation search were similar, since the characteristic
signatures of the processes are nearly identical. Candidate events
were selected based on the following criteria: the event occurred
during the beam gate, had its interaction vertex within the fiducial
volume, and had a single $\mu^-$ reaching the toroid spectrometer.  The
muon was required to be well contained within the toroid and to have
an energy between 15 and 600 GeV.  The muon angle was also required to
be less than 150 mrad with respect to the beam axis.  To reduce the
number of cosmic ray muons entering the selection sample, events which
contained significant activity upstream of the reconstructed vertex were
removed.  The hadronic energy of the interaction was required to be
less than 3 GeV.  Finally, the neutrino beam running mode determined the
sample into which the events were placed.  For IMD, we required a
$\mu^-$ in neutrino mode; for LNV candidates, a $\mu^-$ in
anti-neutrino mode.  For the LNV sample, we also placed an additional
requirement on the $\chi^2$ of the muon track within the toroid, in
order to minimize events where the charge of the muon was
misidentified.

\begin{figure}
\centerline{
\psfig{figure=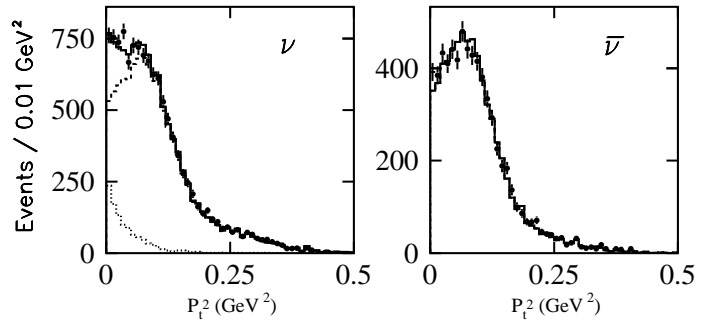,bbllx=0bp,bblly=270bp,bburx=540bp,bbury=525bp,width=3.8in}}
\caption{Transverse momentum distributions for data (crosses) and
Monte Carlo (solid) for right-sign neutrino events (left) and right-sign anti-neutrino
events (right). The plot on the left is broken down into background
only (dashed) and IMD signal (dotted). The plot on the right shows
Monte Carlo background only.}
\label{fig:imd_pt}
\end{figure}

\begin{figure}
\centerline{
\psfig{figure=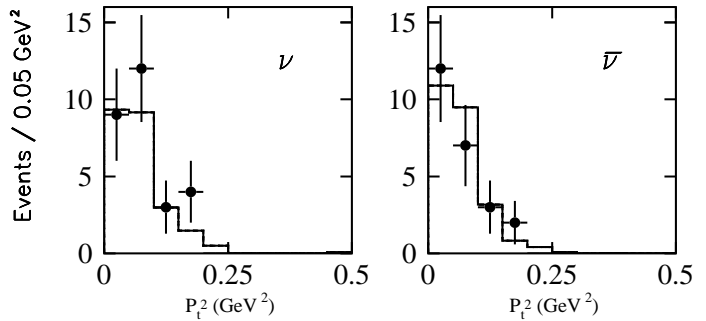,bbllx=0bp,bblly=270bp,bburx=540bp,bbury=525bp,width=3.8in}}
\caption{Transverse momentum distributions for data (crosses) and
Monte Carlo (solid) for neutrino events (left) and anti-neutrino
events (right) with a wrong-sign muon.  A LNV signal would appear as
an excess of events in anti-neutrino mode.}
\label{fig:lnv_pt}
\end{figure}

Because both IMD and LNV events involve neutrino scattering from an
electron, there exists a kinematic limit on the transverse momentum of
the muon: $p_t^2 \leq 2 m_e E_{\nu}$, where $m_e$ is the mass of the
electron and $E_{\nu}$ is the neutrino energy.  We therefore apply an
energy-dependent requirement on the transverse momentum of the event
in order to further isolate signal events.  The cut requires $p^2_t
\leq p_t^{2~max}$, where $p_t^{2~max} \equiv (0.059 + E_{\mu}/671)$
GeV$^2$. This cut, which is based on Monte Carlo signal studies, was
designed to retain 90\% of the signal.  The efficiency after all cuts for IMD
events was 79.6\%. The final efficiency for LNV events was 89.8\% for
V$-$A couplings and 79.9\% for scalar couplings.
Figures~\ref{fig:imd_pt} and \ref{fig:lnv_pt} show the $p^2_t$
distributions for right-sign and wrong-sign events respectively.
Right-sign events are neutrino (anti-neutrino) events with an outgoing
$\mu^-$ $(\mu^+)$; wrong-sign events are the opposite: neutrino
(anti-neutrino) events with a outgoing $\mu^+$ $(\mu^-)$.

For right-sign events, the primary backgrounds that enter the IMD
sample come from low hadronic energy neutrino charged-current
interactions in the detector.  These include quasi-elastic events,
resonance events, and some small fraction of deep-inelastic scattering
(DIS) events with a very small momentum transfer \cite{Smith,Rein}.
Understanding the background levels is essential for the IMD
measurement, since the signal to background ratio for these events is one to eight.  To accomplish this, we perform a full Monte
Carlo simulation of low hadronic energy neutrino processes. To
simulate neutrino resonance production, we used a low-$Q^2$
higher-twist approximation\cite{high_twist}. We found this method more
accurate in averaging over all low-multiplicity states than the
single-pion production model from Rein and Sehgal\cite{Sehgal}.
Nuclear effects such as Fermi motion\cite{Fermi} and Pauli
suppression\cite{Smith,Pauli} were also applied to the Monte Carlo
simulations.  The Monte Carlo was absolutely normalized to data DIS
events with hadronic energies above 30 GeV for each running mode. The
normalization sample contained 0.83(0.25) million neutrino
(anti-neutrino) interactions with a mean energy of 140(120) GeV.

The dominant systematic uncertainties for the right-sign events are
related to the modeling of these low hadronic energy processes.
Systematic errors include effects from muon energy and angular
resolution, background cross-section uncertainties, Pauli suppression,
and MC normalization.  In addition, we take into account
radiative correction errors which affect the IMD cross-section. A
complete list of systematic errors is shown in Table~\ref{tab:sys}.

\begin{table}
\caption{Errors on IMD expected signal and LNV expected background.
Total statistical and systematic errors reflect errors from full
parameter fit, which take into account correlations between errors.}
\label{tab:sys}
\begin{center}
\begin{tabular}{ c c c }
Category & IMD (\%) & LNV (\%) \\
\hline
Statistical Error & $ \pm 6.7$ & $ \pm 13.0$ \\
\hline
Muon Energy Scale & $\pm 1.0 $ & $\pm 2.2$ \\
Hadron Energy Scale & $\pm 0.3 $ & $\pm 0.7$ \\
Angle Smearing & $\pm 0.6 $ & $\pm 1.4 $\\
Normalization & $\pm 6.4 $ & $\pm 2.6$ \\
Quasi-Elastic Cross-Section & $\pm 1.0$ & $ \pm 0.6$ \\
Pauli Suppression & $\pm 8.5$ & $\pm 2.2 $ \\
Beam Impurities & N/A & $\pm 0.7$ \\
Charge Identity & $\ll 0.1$ & $\pm 1.5$ \\
Radiative Corrections & $\pm 1.0$ & $\pm 1.0$ \\
\hline 
Total Systematics (fit) & $\pm 8.2 $ & $\pm 4.4$ \\
\hline
\end{tabular}
\end{center}
\end{table}

The validity of the background modeling was checked directly against
the data by looking at the right-sign, anti-neutrino process
$\overline{\nu}_{\mu} + N \rightarrow \mu^+ + N'$.  This particular
configuration selects only background events, and thus is an ideal
platform to test the data to Monte Carlo agreement and systematics.  A
fit to the anti-neutrino $p^2_t$ distribution (Fig.~\ref{fig:imd_pt})
is performed where the backgrounds are allowed to vary within the
uncertainties shown in the first column of Table~\ref{tab:sys}.  The
fit gives an excellent $\chi^2$/d.o.f. of 44.9/50, indicating that the
background estimate agrees well with the anti-neutrino data within the
systematic uncertainties.

Having verified the size and spectrum of the background, a fit to the
neutrino data is performed to extract the IMD signal. The fit includes
the previously mentioned backgrounds plus an IMD signal contribution
with the proper $p^2_t$ distribution. As before, the backgrounds are
allowed to vary within the uncertainties shown in the first column of
Table~\ref{tab:sys}.  As shown in Fig.~\ref{fig:imd_pt}, the data are
well described by the combination of an IMD signal at low $p^2_t$ plus
the background.  From the fit, we extract a total of $1050 \pm 139$
IMD events, where 1311 events were expected based on Standard Model
predictions, taking into account radiative corrections \cite{Bardin} (see Table~\ref{totalbg}).

\begin{table}
\caption{Signal extraction from Monte Carlo background.}
\label{totalbg}
\vspace{0.4cm}
\begin{center}
\begin{tabular}{l c c r }
\hline
Type & $\nu$ Mode / $\mu$ Charge & Data & Fit Results \\
\hline
IMD Signal & $\nu$ / $\mu^-$ & 11792 & $1050 \pm 139 $ \\
LNV Signal (V$-$A) & $\overline{\nu}$ / $\mu^-$ & 24 & $ 0.6 \pm 3.3$ \\
LNV Signal (scalar) & $\overline{\nu}$ / $\mu^-$ & 24 & $ -0.6 \pm 3.3$ \\
\hline
\end{tabular}
\end{center}
\end{table}

The differential cross-section for IMD can be written as:

\begin{equation}
\frac{d\sigma}{dy} = \sigma_0 \cdot E_{\nu} \cdot (1-r)
\end{equation}

\noindent where $y = E_{\mu}/E_{\nu}$, $r = m_{\mu}^2/s$, $\sigma_0 = \frac{2 m_e G_F^2}{\pi}$, and $s$ is the center-of-mass energy of the interaction. For inverse muon decay, the NuTeV measurement for the IMD
asymptotic cross-section ($E_{\nu} \gg m_{\mu}$) is:

\begin{equation}
\sigma_0 = (13.8 \pm 1.2 \pm 1.4) \times 10^{-42} {\rm~cm}^2/{\rm GeV}
\end{equation}

\noindent where the first error is statistical, and the second is
systematic. The average neutrino energy for the IMD events sampled in the NuTeV experiment is 130
GeV. This measurement is in agreement with the theoretically predicted
value of $17.2 \times 10^{-42}$ cm$^2$/GeV and is also consistent with
the CHARM II measurement of $(16.5 \pm 0.9) \times 10^{-42}$
cm$^2$/GeV \cite{Charm}.

By requiring that the muon charge not match the neutrino running mode
(wrong-sign events), the analysis immediately becomes sensitive to
lepton number violation.  The dominant backgrounds in this case
arise from beam impurities and muon charge mis-identification.  Beam
impurities come mainly from charmed meson decays and decays of wrong-sign hadrons produced in secondary interactions\cite{Drew}. Beam impurities constitute about 72\% of the total LNV
background. Charge mis-identification backgrounds are often associated
with $\delta$-ray production or multiple scattering of the muon in the
toroid spectrometer.  These backgrounds can be greatly reduced by imposing
quality cuts on the muon track in the toroid spectrometer.  The total
fraction of charge mis-identification is 0.06\% for anti-neutrino
running mode. This source contributes 14\% of the LNV
background. Finally, there exists an irreducible background from
$\overline{\nu}_e e^- \rightarrow \mu^- \overline{\nu}_{\mu}$, which
contributes about 14\% of the LNV background.

The generic expression for the
differential LNV cross-section is given by:

\begin{equation}
\frac{d\sigma}{dy} = \lambda \frac{G^2_F s}{\pi} [A_V \cdot y(y-r) +
A_S \cdot (1-r)]
\end{equation}

\noindent where $\lambda$ represents the strength of the interaction
and $A_V$ and $A_S$ determine whether the reaction is V$-$A or
scalar. Integrating over all allowed values of $y$, and normalizing to
the Standard Model IMD cross-section, allows the LNV cross-section to
be written as:

\begin{equation}
\frac{\sigma (\overline{\nu}_{\mu} e^- \rightarrow \mu^-
\overline{\nu}_e)}{\sigma (\nu_{\mu} e^- \rightarrow \mu^- \nu_e)} =
\lambda \cdot [A_V \cdot (\frac{1+r/2}{3}) + A_S]
\end{equation}

We can make a consistency check on the background estimation by
looking at the neutrino process $\nu_{\mu} + N \rightarrow \mu^+ +
N'$.  Momentum distributions of these events are shown in
Fig.~\ref{fig:lnv_pt}.  A total of 28 data events were seen in this
sample where $23.5 \pm 3.7$ (stat. + sys.) were expected, consistent
with the background estimate.

\begin{figure}
\centerline{
\psfig{figure=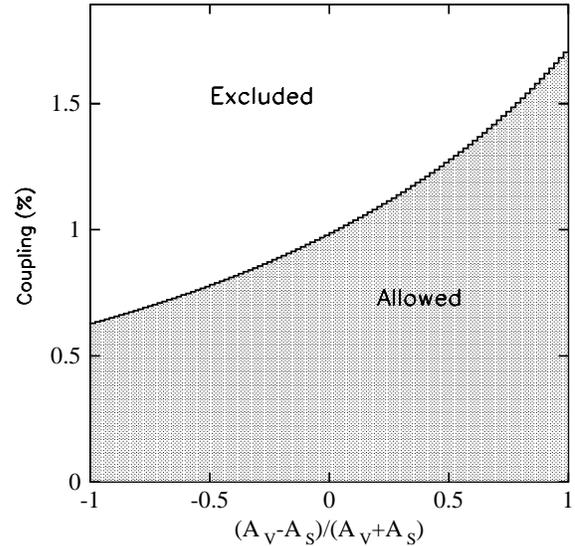,bbllx=0bp,bblly=100bp,bburx=600bp,bbury=600bp,height=3.in}}
\caption{Limit on the lepton number violation process
$\overline{\nu}_{\mu} e^- \rightarrow \mu^- \overline{\nu}_e$ as a
function of the scalar and vector couplings $(A_V-A_S)/(A_V+A_S)$.
The non-shaded region is excluded by this result.}
\label{fig:limits}
\end{figure}

Looking in the LNV signal channel $\overline{\nu}_{\mu} + e^-
\rightarrow \mu^- + \overline{\nu}_e $ yields a total of 24 data
events.  A fit of the $p^2_t$ distribution to only background
sources yields an acceptable $\chi^2$/d.o.f. of 2.5/5, showing
no indication of a LNV signal.  Including a possible LNV signal in the
fit yields a total LNV contribution of $0.6 \pm 3.1 \pm 1.1$
events for a V$-$A coupling and $-0.6 \pm 3.1 \pm 1.1$ events for a
scalar coupling.  As shown in Table~\ref{tab:sys}, the LNV analysis is
dominated by statistical uncertainty.  These fit results can be recast in the form of 90\% C.L. limits on the LNV cross-section as a
function of $(A_V-A_S)/(A_V+A_S)$, as shown in
Figure~\ref{fig:limits}.  If we assume a pure V$-$A coupling, this
yields $\lambda \leq 1.7\%$ while a scalar coupling yields a limit of
$\lambda \leq 0.6\%$.

This limit is currently the most stringent limit obtained directly
from neutrino-electron scattering.  Previous results limited the pure
V$-$A coupling to below 5\%\cite{Charm_lnv}.  The LAMPF experiment has
set an earlier limit from muon decay rates at $\leq
1.2\%$\cite{Los_Alamos} for pure V$-$A couplings.

In conclusion, NuTeV has performed a measurement of the inverse muon
decay cross-section and a direct search for lepton number violation.  The IMD
asymptotic cross-section is measured to be $(13.8 \pm 1.2 \pm 1.4)
\times 10^{-42}$ cm$^2$/GeV. The LNV search limits the strength of the
interaction to be $\leq 1.7\%$ for V$-$A and $\leq 0.6\%$ for scalar
couplings.

This research was supported by the U.S. Department of Energy and the
National Science Foundation.  We thank the staff of FNAL for their
contributions to the construction and support of this experiment
during the 1996-1997 fixed target run.


\begin{references}
\vspace{-.50in} 
%
\bibitem{Fetcher}
W. Fetscher, H.~J.~Gerber, and K.~F.~Johnson, \pl{B173}{102}{1986}
%
\bibitem{Cabbibo}
N. Cabbibo and R. Gatto, \prl{6}{381}{1961}
%
\bibitem{Herczeg}
P. Herczeg and R. N. Mohapatra, \prl{69}{2475}{1992}
%
\bibitem{Cuypers}
F. Cuypers and S. Davidson, \epj{C2}{503}{1998}
%
\bibitem{SSQT} 
J. Yu {\it et al.}, ``Technical Memorandum: NuTeV SSQT
performance,'' Report No. FERMILAB-TM-2040, 1998.
%
\bibitem{nc_charm}
A. Alton {\it et al.}, \prev{D63}{012001}{2001}
%
\bibitem{detector} 
W. Sakumoto {\it et al.}, \nim{A294}{179}{1990}
B. King {\it et al.}, \nim{A302}{254}{1991}
%
\bibitem{NIM}
D. A. Harris, J. Yu, {\it et al.}, \nim{A447}{373}{2000}
%
\bibitem{Smith}
C.H. Llewellyn Smith, \prep{3C}{261}{1971}
%
\bibitem{Rein}
R. Belusevic and D. Rein, \prev{D46}{3747}{1992}
%
\bibitem{high_twist}
U.~K.~Yang and A. Bodek,\prl{82}{2467}{1999}
%
\bibitem{Sehgal}
D. Rein and L. M. Sehgal, Annals of Phys. {\bf 133}, 79 (1981).
%
\bibitem{Fermi}
A. Bodek and J.~L.~Ritchie, \prev{D23}{1070}{1981}
%
\bibitem{Pauli}
E. A. Paschos, L. Pasquali, and J. Y. Yu, \prev{B588}{263}{2000}
%
\bibitem{Bardin}
D.~Yu. Bardin and V.~A.~Dokuchaeva, \np{B287}{839}{1987}
%
\bibitem{Charm}
P. Vilain {\it et al.}, \pl{B364}{121}{1995}
%
\bibitem{Drew} A.~Alton, Ph.D. thesis, Kansas State University,
(2001); A.~Alton {\it et al.}, ``Observation of Neutral Current Charm
Production in $\nu_{\mu}$Fe Scattering at the Tevatron,'' to be published
in Phys. Rev. D (hep-ex/0008068).
%
\bibitem{Charm_lnv}
F. Bergsma {\it et al.}, \pl{122B}{465}{1983}
%
\bibitem{Los_Alamos}
S.~J.~Freedman {\it et al.}, \prev{D47}{811}{1993}
%
\end{references}
\end{document}